\documentclass[11pt,a4paper]{article}
\usepackage{times}
\usepackage[T1]{fontenc}
\usepackage[latin1]{inputenc}
\usepackage[francais]{babel}
\usepackage{fancyhdr}
\usepackage{amssymb}
\usepackage{undertilde}
\usepackage{amsmath}
\usepackage{setspace}
\usepackage{titlesec}
\titleformat{\section}
  {\normalfont\fontsize{16}{20}\bfseries}{\thesection}{1em}{}
\titleformat{\subsection}
  {\normalfont\fontsize{16}{20}\bfseries}{\thesubsection}{1em}{}
\titlespacing*{\section}
{0pt}{2.ex plus 1ex minus .2ex}{.0ex}
\titlespacing*{\subsection}
{0pt}{0.ex}{.0ex}

\begin{document}

\begin{center}
\begin{spacing}{2.05}
{\fontsize{20}{20}
\bf
Equations de conservation et lois de comportement d'un polymËre Èlectroactif\\
}
\end{spacing}
\end{center}
\vspace{-1.25cm}
\begin{center}
{\fontsize{14}{20}
\bf
M. TIXIER\textsuperscript{a}, J. POUGET\textsuperscript{b}\\
\bigskip
}
{\fontsize{12}{20}
a. DÈpartement de Physique, UniversitÈ de Versailles Saint Quentin,
45, avenue des Etats-Unis, F-78035 Versailles, France;
mireille.tixier@uvsq.fr\\
b. Institut Jean le Rond d'Alembert, UMR\ 7190, UniversitÈ Pierre et
Marie Curie, CNRS, F-75005 Paris, France; pouget@lmm.jussieu.fr\\}
\end{center}

\vspace{10pt}

{\fontsize{16}{20}
\bf
R\'esum\'e :
}
\medskip

\textit{Un polym\`{e}re \'{e}lectro-actif ionique (le Nafion par exemple) peut \^{e}%
tre utilis\'{e} comme capteur ou comme actionneur. Pour ce faire, on place
une fine couche de ce mat\'{e}riau satur\'{e} d'eau entre deux \'{e}%
lectrodes. La saturation en eau entra\^{\i}ne une dissociation quasi compl%
\`{e}te du polym\`{e}re et la lib\'{e}ration de cations de petite taille.
L'application d'un champ \'{e}lectrique perpendiculairement \`{a} la lame
provoque la flexion de celle-ci. Inversement, le fl\'{e}chissement de la
lame fait appara\^{\i}tre une diff\'{e}rence de potentiel entre les \'{e}%
lectrodes. Ce ph\'{e}nom\`{e}ne fait intervenir des couplages multiphysiques
de type \'{e}lectro-m\'{e}cano-chimiques.
Nous avons mod\'{e}lis\'{e} ce syst\`{e}me par un milieu poreux d\'{e}%
formable dans lequel s'\'{e}coule une solution ionique et nous avons utilis\'{e} %
une approche "milieu continu". Les \'{e}quations de Maxwell et de conservation de la %
masse, de la quantit\'{e} de mouvement et de l'\'{e}nergie sont \'{e}crites d'abord \`{a} %
l'\'{e}chelle microscopique pour chaque phase et pour les interfaces, puis pour le
mat\'{e}riau complet gr\^{a}ce \`{a} une technique de moyenne. La
thermodynamique des processus irr\'{e}versibles lin\'{e}aires nous permet
d'en d\'{e}duire les lois de comportement : une loi rh\'{e}ologique de type
Kelvin-Voigt, des lois de Fourier et de Darcy g\'{e}n\'{e}ralis\'{e}es et
une \'{e}quation de type Nernst-Planck.}

\vspace{20pt}

{\fontsize{16}{20}
\bf
Abstract :
}
\bigskip

\textit{Ionic electro-active polymer (Nafion for example) can be used as sensor or
actuator. To this end, a thin film of the water-saturated material is
sandwiched between two electrodes. Water saturation causes a quasi-complete
dissociation of the polymer and the release of small cations. The
application of an electric field across the thickness results in the bending
of the strip. Conversely, a voltage can be detected between the two
electrodes when the strip is bent. This phenomenon involves multiphysics
couplings of electro-mechanical-chemical type.
 The system is modeled by a deformable porous medium in which flows an
ionic solution and we use a continuous
medium approach. Maxwell's equations and conservation laws of mass, linear
momentum and energy are first written at the microscopic scale for each
phase and interfaces, then for the complete material using an average
technique. Thermodynamics of linear irreversible processes provides the
constitutive equations : a Kelvin-Voigt stress-strain relation, generalized
Fourier's and Darcy's laws and a Nernst-Planck equation.}

\vspace{28pt}

{\fontsize{14}{20}
\bf
Mots clefs :  Electro-active polymers - Multiphysics coupling - Deformable porous
media - Balance laws - Constitutive relations - Polymer mechanics - Nafion
}

\section{Introduction}
\medskip
Les polym\`{e}res \'{e}lectro-actifs sont des mat\'{e}riaux innovants tr\`{e}%
s int\'{e}ressants, notamment dans les domaines de la bio-inspiration
(conception d'ailes battantes inspir\'{e}es du vol des insectes), de la biom%
\'{e}canique (muscles artificiels) et de la r\'{e}cup\'{e}ration d'\'{e}%
nergie. Nous nous sommes plus particuli\`{e}rement int\'{e}ress\'{e}s \`{a}
une lame de poly\'{e}lectrolyte de type Nafion recouverte sur ses deux faces
d'une fine couche de m\'{e}tal servant d'\'{e}lectrodes (I.P.M.C.). Un tel
syst\`{e}me pr\'{e}sente des d\'{e}formations de grande amplitude lorsqu'il est
soumis \`{a} des diff\'{e}rences de potentiel de quelques volts;
inversement, la flexion de la lame fait appara\^{\i}tre une diff\'{e}rence
de potentiel entre les \'{e}lectrodes. Il peut donc \^{e}tre utilis\'{e}
comme capteur ou actionneur.

Lorsque le poly\'{e}lectrolyte est satur\'{e} d'eau, il se dissocie quasi
compl\`{e}tement, lib\'{e}rant des cations de petite taille ($H^{+}$, $%
Li^{+} $ ou $Na^{+}$); les anions restent fix\'{e}s sur le squelette du polym%
\`{e}re. Lorsqu'un champ \'{e}lectrique est appliqu\'{e} perpendiculairement
aux \'{e}lectrodes, les cations se d\'{e}placent vers l'\'{e}lectrode n\'{e}%
gative (cathode), entra\^{\i}nant avec eux le solvant par un ph\'{e}nom\`{e}%
ne d'osmose. Ceci provoque un gonflement du polym\`{e}re au voisinage de la
cathode et une diminution de volume du c\^{o}t\'{e} oppos\'{e}. Il en r\'{e}%
sulte un fl\'{e}chissement de la lame vers l'anode. Une lame de $200\;\mu m$
d'\'{e}paisseur de quelques centim\`{e}tres de long fl\'{e}chit ainsi de
quelques millim\`{e}tres en une seconde sous l'action d'une diff\'{e}rence
de potentiel de quelques volts \cite{nemat2000}. La mod\'{e}lisation de ce
mat\'{e}riau doit donc prendre en compte les couplages entre ph\'{e}nom\`{e}%
nes \'{e}lectriques, chimiques et m\'{e}caniques.

\section{ModÈlisation}
\medskip
Notre mod\'{e}lisation est bas\'{e}e sur la thermom\'{e}canique des milieux
continus. Les cha\^{\i}nes polym\`{e}res charg\'{e}es n\'{e}gativement sont
assimil\'{e}es \`{a} un milieu poreux d\'{e}formable, homog\`{e}ne et
isotrope. Ce milieu poreux est satur\'{e} d'une solution ionique constitu%
\'{e}e par l'eau et les cations. Le mat\'{e}riau appara\^{\i}t donc comme la
superposition de trois syst\`{e}mes mobiles les uns par rapport aux autres :
les cations, le solvant et le solide poreux. Les grandeurs physiques
relatives \`{a} ces trois syst\`{e}mes sont identifi\'{e}es respectivement
par les indices $1$, $2$ et $3$, l'indice $4$ \'{e}tant relatif \`{a} la
phase liquide (eau + cations) et l'absence d'indice au mat\'{e}riau complet.
Les phases solide et liquide sont s\'{e}par\'{e}es par une interface d'\'{e}%
paisseur n\'{e}gligeable (indice $i$). Les constituants $2$, $3$ et $4~$sont
assimil\'{e}s \`{a} des milieux continus, de m\^{e}me que le mat\'{e}riau
complet. Nous supposerons en outre que la gravit\'{e} et le champ et l'induction magn\'{e}%
tiques sont n\'{e}gligeables.

\subsection{Processus de moyenne}
\medskip
Nous avons utilis\'{e} un mod\`{e}le \`{a} gros grains d\'{e}velopp\'{e}
pour les m\'{e}langes \`{a} deux constituants \cite{Nigmatulin79,Drew83,Ishii06,Lhuilier03}. %
Les diff\'{e}rentes grandeurs
physiques sont tout d'abord d\'{e}finies \`{a} l'\'{e}chelle microscopique.
Cette \'{e}chelle doit \^{e}tre suffisamment petite pour que le volume
correspondant ne contienne qu'une seule phase, mais suffisamment grande pour
l\'{e}gitimer l'hypoth\`{e}se de milieu continu. A l'\'{e}chelle
macroscopique, on d\'{e}finit un volume \'{e}l\'{e}mentaire repr\'{e}%
sentatif (V.E.R.) contenant les deux phases; ce volume doit \^{e}tre
suffisamment grand pour que les grandeurs relatives au mat\'{e}riau complet
aient un sens, mais suffisamment petit pour que l'on puisse les consid\'{e}%
rer comme locales. Dans le cas du Nafion, l'\'{e}chelle microscopique est de
l'ordre d'une centaine d'Angstr\"{o}ms et l'\'{e}chelle macroscopique de
l'ordre du micron \cite{Gierke}. Pour chacune des phases 3 et 4, on d\'{e}%
finit une fonction de pr\'{e}sence $\chi _{k}\left( \overrightarrow{r}%
,t\right) $ de type Heaviside :%
\begin{equation}
\chi _{k}=1\;si\;la\;phase\;k\;occupe\;le\;point\;\overrightarrow{r}%
\;au\;temps\;t,\quad \chi _{k}=0\; sinon
\end{equation}%
Les grandeurs physiques microscopiques sont rep\'{e}r\'{e}es par un exposant 
$^{0}$, les grandeurs macroscopiques sont sans exposant. Une grandeur
physique macroscopique $g_{k}$ est obtenue en calculant la moyenne
statistique $\left\langle {}\right\rangle _{k}$ sur le V.E.R. d'une grandeur
microscopique $g_{k}^{0}$ relative \`{a} la phase $k$. Nous supposons que
cette moyenne est \'{e}quivalente \`{a} une moyenne en volume (hypoth\`{e}se
d'ergodicit\'{e}) et qu'elle commute avec les d\'{e}riv\'{e}es spatiales et
temporelles \cite{Drew83,Lhuilier03} : 
\begin{equation}
g_{k}=\left\langle \chi _{k}g_{k}^{0}\right\rangle =\phi _{k}\left\langle
g_{k}^{0}\right\rangle _{k}
\end{equation}%
o\`{u} $\phi _{k}=\left\langle \chi _{k}\right\rangle $ d\'{e}signe la
fraction volumique de la phase $k$. On remarquera que les grandeurs
macroscopiques sont d\'{e}finies sur tout le mat\'{e}riau, alors que les
grandeurs microscopiques ne sont d\'{e}finies que sur une phase. Une
grandeur macroscopique $g$ relative au mat\'{e}riau complet est obtenue par
sommation des grandeurs macroscopiques relatives aux diff\'{e}rentes phases
et interfaces :%
\begin{equation}
g=\sum\limits_{k=3,4,i}g_{k}
\end{equation}

\subsection{ModÈlisation de l'interface}
\medskip
Dans la r\'{e}alit\'{e}, la zone de contact entre les deux phases a une
certaine \'{e}paisseur et les grandeurs physiques extensives varient contin\^{u}%
ment d'une phase \`{a} l'autre. On peut remplacer cette r\'{e}alit%
\'{e} complexe par deux phases volumiques dont les param\`{e}tres
microscopiques peuvent \^{e}tre consid\'{e}r\'{e}s comme localement
constants s\'{e}par\'{e}es par une surface de discontinuit\'{e} $\Sigma $ de
position arbitraire. Soit $\Omega $, un cylindre de bases parall\`{e}les 
\`{a} $\Sigma $ \`{a} cheval sur l'interface. $\Omega $ est donc divis\'{e}
en deux volumes $\Omega _{3}$ et $\Omega _{4}$ appartenant respectivement
aux phases 3 et 4.

Les grandeurs continues d\'{e}crivant la zone de contact seront rep\'{e}r%
\'{e}es par un exposant $^{0}$ et pas d'indice. Une grandeur surfacique
microscopique $g_{i}^{0}$ attach\'{e}e l'interface peut \^{e}tre d\'{e}finie
par :%
\begin{equation}
g_{i}^{0}=\underset{\Sigma \longrightarrow 0}{\lim }\frac{1}{\Sigma }\left\{
\int_{\Omega }g^{0}dv-\int_{\Omega _{3}}g_{3}^{0}dv-\int_{\Omega
_{4}}g_{4}^{0}dv\right\}
\end{equation}%
La grandeur macroscopique correspondante est donn\'{e}e par :%
\begin{equation}
g_{i}=\left\langle g_{i}^{0}\chi _{i}\right\rangle
\end{equation}
o\`{u} $\chi _{i}=-\overrightarrow{grad} \chi _{k}.\overrightarrow{n _{k}}$ d\'{e}signe 
la fonction de pr\'{e}sence de l'interface et $\overrightarrow{n _{k}}$ la normale 
sortante de la phase $k$. La position de $\Sigma $ est fix\'{e}e de telle sorte que l'interface n'ait
pas de masse volumique ($\rho _{i}^{0}=0$). On suppose par ailleurs qu'il
n'y a pas de flux de mati\`{e}re d'une phase dans l'autre, d'o\`{u} :%
\begin{equation}
\overrightarrow{V_{1}^{0}}=\overrightarrow{V_{2}^{0}}=\overrightarrow{%
V_{3}^{0}}=\overrightarrow{V_{4}^{0}}=\overrightarrow{V_{i}^{0}}
\end{equation}%
o\`{u} $\overrightarrow{V_{k}^{0}}$ d\'{e}signe la vitesse microscopique de
la phase $k$. On n\'{e}gligera en outre les fluctuations de vitesses de tous
les constituants et de l'interface \`{a} l'\'{e}chelle du V.E.R.

\subsection{DÈrivÈes particulaires et matÈrielles}
\medskip
Pour \'{e}crire les \'{e}quations de bilan, il est n\'{e}cessaire de
calculer les variations d'une grandeur extensive $g_{k}$ en suivant le mouvement
 de la phase correspondante. C'est ce que
nous appellerons une d\'{e}riv\'{e}e particulaire $\frac{d_{k}}{dt}$ :%
\begin{equation}
\frac{d_{k}g_{k}}{dt}=\frac{\partial g_{k}}{\partial t}+\overrightarrow{grad}%
g_{k}.\overrightarrow{V_{k}}
\end{equation}%
Cette d\'{e}riv\'{e}e peut \^{e}tre d\'{e}finie \`{a} l'\'{e}chelle
microscopique ou macroscopique.

Les diff\'{e}rentes phases ne se d\'{e}pla\c{c}ent pas \`{a} la m\^{e}me
vitesse. Pour calculer la variation d'une grandeur extensive $g$ relative au
mat\'{e}riau complet, nous d\'{e}finissons une d\'{e}riv\'{e}e en suivant le
mouvement des diff\'{e}rents constituants appel\'{e}e d\'{e}riv\'{e}e mat%
\'{e}rielle \cite{Coussy95,Biot77} :%
\begin{equation}
\begin{tabular}{l}
$\rho \frac{D}{Dt}\left( \frac{g}{\rho }\right) =\rho \sum\limits_{k}\frac{%
\rho _{k}}{\rho }\frac{d_{k}}{dt}\left( \frac{g_{k}}{\rho _{k}}\right)
=\sum\limits_{3,4,i}\frac{\partial g_{k}}{\partial t}+div\left( g_{k}%
\overrightarrow{V_{k}}\right) $ \\ 
$\rho \frac{D}{Dt}\left( \frac{\overrightarrow{g}}{\rho }\right)
=\sum\limits_{3,4,i}\frac{\partial \overrightarrow{g_{k}}}{\partial t}%
+div\left( \overrightarrow{g_{k}}\otimes \overrightarrow{V_{k}}\right) $%
\end{tabular}%
\end{equation}%
o\`{u} $\rho _{k}$ d\'{e}signe la masse volumique de la phase $k$ rapport%
\'{e}e au volume du mat\'{e}riau complet. Cette d\'{e}riv\'{e}e n'a bien s%
\^{u}r de sens qu'\`{a} l'\'{e}chelle macroscopique. Elle ne doit pas \^{e}%
tre confondue avec la d\'{e}riv\'{e}e particulaire $\frac{d}{dt}$ en suivant
la vitesse barycentrique $\overrightarrow{V}=\sum\limits_{k=3,4}\frac{\rho
_{k}}{\rho }\overrightarrow{V_{k}}$ du mat\'{e}riau.

\subsection{Equations de bilan}
\medskip
L'\'{e}quation de bilan d'une grandeur extensive microscopique relative \`{a}
la phase $k$ de densit\'{e} volumique $g_{k}^{0}\left( \overrightarrow{x}%
,t\right) $ est de la forme :%
\begin{equation}
\frac{\partial g_{k}^{0}}{\partial t}+div\left( g_{k}^{0}\overrightarrow{%
V_{k}^{0}}\right) =-div\overrightarrow{A_{k}^{0}}+B_{k}^{0}
\end{equation}%
o\`{u} $\overrightarrow{A_{k}^{0}}$ d\'{e}signe le flux de $g_{k}^{0}$ li%
\'{e} \`{a} des ph\'{e}nom\`{e}nes autres que la convection et $B_{k}^{0}$
la production volumique de $g_{k}^{0}$ (terme source). A l'\'{e}chelle
macroscopique, cette \'{e}quation devient :%
\begin{equation}
\frac{\partial g_{k}}{\partial t}+div\left( g_{k}\overrightarrow{V_{k}}%
\right) =-div\overrightarrow{A_{k}}+B_{k}-\left\langle \overrightarrow{%
A_{k}^{0}}.\overrightarrow{n_{k}}\chi _{i}\right\rangle
\end{equation}%
o\`{u} :%
\begin{equation}
\overrightarrow{A_{k}}=\left\langle \chi _{k}\overrightarrow{A_{k}^{0}}%
\right\rangle \qquad \qquad \qquad B_{k}=\left\langle \chi
_{k}B_{k}^{0}\right\rangle
\end{equation}

Pour une interface, l'\'{e}quation de bilan s'\'{e}crit \cite{Ishii06} :%
\begin{equation}
\frac{\partial g_{i}^{0}}{\partial t}+div_{s}\left( g_{i}^{0}\overrightarrow{%
V_{i}^{0}}\right) =\sum\limits_{3,4}\left[ g_{k}^{0}\left( \overrightarrow{%
V_{k}}-\overrightarrow{V_{i}^{0}}\right) .\overrightarrow{n_{k}}+%
\overrightarrow{A_{k}^{0}}.\overrightarrow{n_{k}}\right] -div_{s}%
\overrightarrow{A_{i}^{0}}+B_{i}^{0}
\end{equation}%
o\`{u} $g_{i}^{0}$ est une grandeur surfacique. $div_{s}$ d\'{e}signe la
divergence le long de l'interface, $\overrightarrow{A_{i}^{0}}$ est le flux
de $g_{i}^{0}$ le long de l'interface li\'{e} \`{a} des ph\'{e}nom\`{e}nes
autres que la convection et $B_{i}^{0}$ la production surfacique de $%
g_{i}^{0}$. A l'\'{e}chelle macroscopique : 
\begin{equation}
\frac{\partial g_{i}}{\partial t}+div\left( g_{i}\overrightarrow{V_{i}}%
\right) =\sum\limits_{3,4}\left\langle \chi _{i}\overrightarrow{A_{k}^{0}}.%
\overrightarrow{n_{k}}\right\rangle -div\overrightarrow{A_{i}}+B_{i}
\end{equation}%
On notera que $g_{i}$, $\overrightarrow{A_{i}}$ et $B_{i}$ sont des
grandeurs volumiques.

Pour le mat\'{e}riau complet, on obtient par sommation :%
\begin{equation}
\rho \frac{D}{Dt}\left( \frac{g}{\rho }\right) =-div\overrightarrow{A}+B
\end{equation}%
o\`{u} :%
\begin{equation}
\overrightarrow{A}=\sum\limits_{3,4,i}\overrightarrow{A_{k}}\qquad \qquad
\qquad B=\sum\limits_{3,4,i}B_{k}
\end{equation}

\section{Equations de conservation}

\subsection{Conservation de la masse}
\medskip
L'obtention de ces \'{e}quations est d\'{e}taill\'{e}e dans \cite{Tixier}. 
La conservation de la masse s'\'{e}crit pour chacune des
phases et pour le mat\'{e}riau complet\ :%
\begin{equation}
\frac{\partial \rho _{k}}{\partial t}+div\left( \rho _{k}\overrightarrow{%
V_{k}}\right) =0\qquad (k=2,3) \qquad \qquad  
\frac{\partial \rho }{\partial t}+div\left( \rho \overrightarrow{V}\right)=0%
\end{equation}%
avec :%
\begin{equation}
\rho _{1}=\phi _{4}CM_{1}\qquad \qquad \rho _{k}=\phi _{k}\rho
_{k}^{0}\qquad (k=2,3)\qquad \qquad \rho _{4}=\rho _{2}+\phi _{4}CM_{1}
\end{equation}%
$M_{k}$ d\'{e}signe la masse molaire et $C$ la
concentration en cations relative au volume de la solution. On suppose les
fluctuations de $C$ n\'{e}gligeables sur le V.E.R..

\subsection{Equations Èlectriques}
\medskip
L'\'{e}quation de conservation de la charge \'{e}lectrique s'\'{e}crit\ :%
\begin{equation}
div\overrightarrow{I}+\frac{\partial \rho Z}{\partial t}=0
\end{equation}%
o\`{u} :%
\begin{equation}
\rho Z=\sum\limits_{3,4}\rho _{k}Z_{k}+Z_{i}\qquad \qquad \overrightarrow{%
I_{3}}=\rho _{3}Z_{3}\overrightarrow{V_{3}}\qquad \qquad \overrightarrow{%
I_{4}}=\rho _{1}Z_{1}\overrightarrow{V_{1}}
\end{equation}%
$Z_{k}$ d\'{e}signe la charge \'{e}lectrique massique et $\overrightarrow{%
I_{k}}$ la densit\'{e} volumique de courant de la phase $k$; $Z_{i}$ est la
densit\'{e} surfacique de charges de l'interface. Le champ \'{e}lectrique $%
\overrightarrow{E_{k}^{0}}$ et l'induction \'{e}lectrique $\overrightarrow{%
D_{k}^{0}}$ v\'{e}rifient les \'{e}quations de Maxwell et leurs conditions
aux limites. On admet que les fluctuations du champ \'{e}lectrique \`{a}
l'\'{e}chelle du V.E.R. sont n\'{e}gligeables et qu'il a m\^{e}me valeur dans 
toutes les phases. On en dÈduit que le polym\`{e}re se comporte 
comme un milieu di\'{e}lectrique lin\'{e}aire, homog\`{e}ne et isotrope :%
\begin{equation}
\begin{tabular}{lll}
$\overrightarrow{rot}\overrightarrow{E}=\overrightarrow{0}\qquad \qquad $ & $%
div\overrightarrow{D}=\rho Z\qquad \qquad $ & $\overrightarrow{D}%
=\varepsilon \overrightarrow{E}$%
\end{tabular}%
\end{equation}%
o˘ la permittivit\'{e} di\'{e}lectrique $\varepsilon$ de la phase $k$ s'Ècrit :
\begin{equation}
\varepsilon =\sum\limits_{3,4}\phi _{k}\left\langle \varepsilon
_{k}^{0}\right\rangle _{k}
\end{equation}%

\subsection{Bilan de la quantitÈ de mouvement}
\medskip
La seule force volumique appliqu\'{e}e est la force \'{e}lectrique. Pour le
mat\'{e}riau complet :%
\begin{equation}
\rho \frac{D\overrightarrow{V}}{Dt}=\overrightarrow{div}\utilde{\sigma }%
+\rho Z\overrightarrow{E}  \label{CQ}
\end{equation}%
On v\'{e}rifie que $\utilde{\sigma }$, tenseur des contraintes du mat\'{e}%
riau global, est sym\'{e}trique et qu'en l'absence de forces ext\'{e}rieures, 
la quantit\'{e} de mouvement se conserve.

\subsection{Equations de bilan d'Ènergie}
\medskip
Les deux phases peuvent \^{e}tre assimil\'{e}es \`{a} des milieux lin\'{e}%
aires isotropes non dissipatifs. Le th\'{e}or\`{e}me de Poynting prend alors
la forme int\'{e}grale suivante si aucune charge ne sort du volume $\Omega $
de fronti\`{e}re $\partial \Omega $ \cite{Jackson} :%
\begin{equation}
\frac{d}{dt}\int_{\Omega }\frac{1}{2}\left( \overrightarrow{E}\cdot 
\overrightarrow{D}+\overrightarrow{B}\cdot \overrightarrow{H}\right)
dv=-\oint\nolimits_{\partial \Omega }\left( \overrightarrow{E}\wedge 
\overrightarrow{H}\right) \cdot \overrightarrow{n}ds-\int_{\Omega }%
\overrightarrow{E}\cdot \overrightarrow{I}dv
\end{equation}%
Le membre de gauche repr\'{e}sente la variation de l'\'{e}nergie potentielle
du domaine $\Omega $ en suivant le mouvement des charges. Pour le mat\'{e}%
riau complet l'\'{e}nergie potentielle $E_{p}=\frac{1}{2}\overrightarrow{D}%
\cdot \overrightarrow{E}$ v\'{e}rifie :%
\begin{equation}
\rho \frac{D}{Dt}\left( \frac{E_{p}}{\rho }\right) =-\overrightarrow{E}\cdot 
\overrightarrow{I}  \label{Ep}
\end{equation}

Les vitesses relatives des diff\'{e}rentes phases $\overrightarrow{V_{k}}-%
\overrightarrow{V}$ sont petites par rapport aux vitesses $\overrightarrow{%
V_{k}}$ mesur\'{e}es dans le r\'{e}f\'{e}rentiel du laboratoire; dans l'exp%
\'{e}rience \'{e}voqu\'{e}e pr\'{e}c\'{e}demment pour le Nafion, on obtient
ainsi des vitesses relatives de l'ordre de $\mathit{2}\;\mathit{10}%
^{-4}\;m\;s^{-1}$ et des vitesses absolues voisines de $\mathit{4}\;\mathit{%
10}^{-3}\;m\;s^{-1}$. En premi\`{e}re approximation, on peut donc identifier
l'\'{e}nergie cin\'{e}tique du mat\'{e}riau complet $E_{c}=\frac{1}{2}\rho
V^{2}$ et la somme des \'{e}nergies cin\'{e}tiques des constituants $%
\sum\limits_{3,4}\frac{1}{2}\rho _{k}V_{k}^{2}$. L'\'{e}quation de bilan de
l'\'{e}nergie cin\'{e}tique se d\'{e}duit de l'\'{e}quation de bilan de la
quantit\'{e} de mouvement et s'\'{e}crit, pour le mat\'{e}riau complet :%
\begin{equation}
\rho \frac{D}{Dt}\left( \frac{E_{c}}{\rho }\right) =\sum\limits_{3,4}\left[
div\left( \utilde{\sigma}_{k}\cdot \overrightarrow{V_{k}}\right) -%
\utilde{\sigma}_{k}:\utilde{grad}\overrightarrow{V_{k}}\right] +\left( 
\overrightarrow{I}-\overrightarrow{i}\right) \cdot \overrightarrow{E}
\end{equation}%
o\`{u} :%
\begin{equation}
\overrightarrow{i}=\overrightarrow{I}-\sum\limits_{k=3,4}\left( \rho
_{k}Z_{k}\overrightarrow{V_{k}}\right) -Z_{i}\overrightarrow{V_{i}}
\end{equation}
est le courant de diffusion des cations dans la solution.

L'\'{e}quation de conservation de l'\'{e}nergie totale $E$ s'\'{e}crit :%
\begin{equation}
\rho \frac{D}{Dt}\left( \frac{E}{\rho }\right) =div\left( \sum\limits_{k=3,4}%
\utilde{\sigma}_{k}\cdot \overrightarrow{V_{k}}\right) -div\overrightarrow{Q}
\end{equation}%
o\`{u} $\overrightarrow{Q}$ d\'{e}signe le flux de chaleur par conduction.

L'\'{e}nergie interne $U$ s'obtient par diff\'{e}rence entre les \'{e}%
nergies totale, potentielle et cin\'{e}tique :%
\begin{equation}
\rho \frac{D}{Dt}\left( \frac{U}{\rho }\right) =\sum\limits_{3,4}\left( %
\utilde{\sigma}_{k}:\utilde{grad}\overrightarrow{V_{k}}\right) +%
\overrightarrow{i}\cdot \overrightarrow{E}-div\overrightarrow{Q}
\end{equation}%
avec $U=E-E_{c}-E_{p}$. On peut \'{e}galement \'{e}crire cette \'{e}quation
en utilisant la d\'{e}riv\'{e}e en suivant le mouvement du barycentre des
constituants du syst\`{e}me :%
\begin{equation}
\rho \frac{d}{dt}\left( \frac{U}{\rho }\right) =\utilde{\sigma }:%
\utilde{grad}\overrightarrow{V}+\overrightarrow{i^{\prime }}\cdot 
\overrightarrow{E}-div\overrightarrow{Q^{\prime }}  \label{U}
\end{equation}%
o\`{u} :%
\begin{equation}
\begin{tabular}{l}
$\overrightarrow{i^{\prime }}=\overrightarrow{I}-\rho Z\overrightarrow{V}%
\simeq \sum\limits_{k=1,3}\rho _{k}Z_{k}\left( \overrightarrow{V_{k}}-%
\overrightarrow{V}\right) $ \\ 
$\overrightarrow{Q^{\prime }}=\overrightarrow{Q}-\sum\limits_{k=3,4}U_{k}%
\left( \overrightarrow{V}-\overrightarrow{V_{k}}\right) -\sum\limits_{k=3,4}%
\utilde{\sigma _{k}}\cdot \left( \overrightarrow{V_{k}}-\overrightarrow{V}%
\right) $%
\end{tabular}%
\end{equation}

On peut r\'{e}sumer les \'{e}changes d'\'{e}nergies gr\^{a}ce au tableau
suivant :%
\begin{equation}
\begin{tabular}{ccccc}
\hline
& flux & $E_{c}\longleftrightarrow E_{p}$ & $U\longleftrightarrow E_{p}$ & $%
E_{c}\longleftrightarrow U$ \\ \hline
$E_{p}$ &  & $-\left( \overrightarrow{I}-\overrightarrow{i}\right) \cdot 
\overrightarrow{E}$ & $-\overrightarrow{i}\cdot \overrightarrow{E}$ &  \\ 
$E_{c}$ & $\sum\limits_{3,4}\utilde{\sigma}_{k}\cdot 
\overrightarrow{V_{k}} $ & $+\left( \overrightarrow{I}-%
\overrightarrow{i}\right) \cdot \overrightarrow{E}$ &  & $-\sum\limits_{3,4}%
\utilde{\sigma}_{k}:\utilde{grad}\overrightarrow{V_{k}}$ \\ 
$U$ & $ -\overrightarrow{Q} $ &  & $+\overrightarrow{i}\cdot 
\overrightarrow{E}$ & $+\sum\limits_{3,4}\left( \utilde{\sigma}_{k}:%
\utilde{grad}\overrightarrow{V_{k}}\right) $ \\ 
$E$ & $ \sum\limits_{3,4}\utilde{\sigma}_{k}\cdot \overrightarrow{%
V_{k}}-\overrightarrow{Q} $ &  &  &  \\ \hline
\end{tabular}%
\end{equation}%
Le flux d'\'{e}nergie cin\'{e}tique est \'{e}gal au travail des forces de
contact, le flux d'\'{e}nergie interne est le flux de chaleur et le flux d'%
\'{e}nergie totale est la somme des deux; on v\'{e}rifie qu'il n'y a pas de
terme source dans cette derni\`{e}re \'{e}quation.
$\left( \overrightarrow{I}-%
\overrightarrow{i}\right) \cdot \overrightarrow{E}$ est le travail des
forces \'{e}lectriques et correspond \`{a} une conversion d'\'{e}nergie
potentielle en \'{e}nergie cin\'{e}tique. $\overrightarrow{i}\cdot 
\overrightarrow{E}$ repr\'{e}sente l'\'{e}nergie potentielle convertie en
chaleur par effet Joule. $\sum\limits_{3,4}\left( \utilde{\sigma}_{k}:%
\utilde{grad}\overrightarrow{V_{k}}\right) $ traduit la dissipation
visqueuse, c'est \`{a} dire la conversion d'\'{e}nergie cin\'{e}tique en
chaleur.

\subsection{Equation de bilan d'entropie}
\medskip
L'\'{e}quation de bilan de l'entropie volumique $S$ s'\'{e}crit :%
\begin{equation}
\rho \frac{D}{Dt}\left( \frac{S}{\rho }\right) =s-div\overrightarrow{\Sigma }
\end{equation}%
o\`{u} $\overrightarrow{\Sigma }$ et $s$ d\'{e}signent respectivement le
flux et la cr\'{e}ation d'entropie. Dans le r\'{e}f\'{e}rentiel
barycentrique, cette \'{e}quation devient :%
\begin{equation}
\rho \frac{d}{dt}\left( \frac{S}{\rho }\right) =s-div\overrightarrow{\Sigma
^{\prime }}  \label{S}
\end{equation}%
o\`{u} :%
\begin{equation}
\overrightarrow{\Sigma ^{\prime }}=\overrightarrow{\Sigma }%
-\sum\limits_{k=3,4}S_{k}\left( \overrightarrow{V}-\overrightarrow{V_{k}}%
\right)
\end{equation}

\section{Fonction de dissipation}

\subsection{Relations thermodynamiques}
\medskip
La relation de Gibbs s'\'{e}crit pour les phases solide et liquide \cite{De Groot} :%
\begin{equation}
\begin{tabular}{l}
$\rho _{3}^{0}\frac{d_{3}^{0}}{dt}\left( \frac{U_{3}^{0}}{\rho _{3}^{0}}%
\right) =p_{3}^{0}\frac{1}{\rho _{3}^{0}}\frac{d_{3}^{0}\rho _{3}^{0}}{dt}+%
\utilde{\sigma _{3}^{0e}}^{s}:\frac{d_{3}^{0}\utilde{\varepsilon _{3}^{0}}%
^{s}}{dt}+\rho _{3}^{0}T_{3}^{0}\frac{d_{3}^{0}}{dt}\left( \frac{S_{3}^{0}}{%
\rho _{3}^{0}}\right) $ \\ 
$\frac{d_{4}^{0}}{dt}\left( \frac{U_{4}^{0}}{\rho _{4}^{0}}\right) =T_{4}^{0}%
\frac{d_{4}^{0}}{dt}\left( \frac{S_{4}^{0}}{\rho _{4}^{0}}\right) -p_{4}^{0}%
\frac{d_{4}^{0}}{dt}\left( \frac{1}{\rho _{4}^{0}}\right)
+\sum\limits_{k=1,2}\mu _{k}^{0}\frac{d_{4}^{0}}{dt}\left( \frac{\rho
_{k}^{\prime }}{\rho _{4}^{0}}\right) $%
\end{tabular}%
\end{equation}%
o\`{u} $T_{k}^{0}$ d\'{e}signe la temp\'{e}rature absolue, $%
\utilde{\varepsilon _{3}^{0}}$ et $\utilde{\sigma _{3}^{0e}}$ les tenseurs 
des d\'{e}formations et des contraintes d'\'{e}quilibre du solide. 
$\utilde{\varepsilon _{3}^{0s}}$ et $\utilde{\sigma _{3}^{0es}}$ %
sont les parties sym\'{e}triques de traces nulles de ces m\^{e}mes tenseurs. 
La pression $p_{3}^{0}$ du solide est d%
\'{e}finie \`{a} partir du tenseur des contraintes et v\'{e}rifie la
relation d'Euler, de m\^{e}me que celle du liquide :%
\begin{equation}
\begin{tabular}{l}
$p_{3}^{0}=-\frac{1}{3}tr\left( \utilde{\sigma_{3}^{0e}}\right)
=T_{3}^{0}S_{3}^{0}-U_{3}^{0}+\mu _{3}^{0}\rho _{3}^{0}$ \\ 
$U_{4}^{0}-T_{4}^{0}S_{4}^{0}+p_{4}^{0}=\sum\limits_{k=1,2}\mu _{k}^{0}\rho
_{k}^{\prime }$%
\end{tabular}%
\end{equation}
o\`{u} $\mu _{k}^{0}$ est le potentiel chimique massique du constituant $k$. 
$\rho _{k}^{\prime }$ d\'{e}signe la masse volumique des constituants rapport%
\'{e}e au volume de la solution :%
\begin{equation}
\begin{tabular}{lll}
$\frac{\rho _{k}^{\prime }}{\rho _{4}^{0}}=\frac{\rho _{k}}{\rho _{4}}\qquad 
$ & $\rho _{1}^{\prime }=CM_{1}\qquad $ & $\rho _{2}^{\prime }=\frac{\rho
_{2}^{0}\phi _{2}}{\phi _{4}}$%
\end{tabular}%
\end{equation}

On peut raisonnablement supposer que les fluctuations des grandeurs
intensives (pressions, temp\'{e}ratures, potentiels chimiques), des tenseurs 
des dÈformations $\utilde{\varepsilon _{3}^{0}}$ et des contraintes d'%
\'{e}quilibre $\utilde{\sigma _{3}^{0e}}$ sont n\'{e}gligeables \`{a} l'\'{e}%
chelle du V.E.R.. Si l'on fait de plus l'hypoth\`{e}se de l'\'{e}quilibre
thermodynamique local, il vient :%
\begin{equation}
\begin{tabular}{l}
$p=p_{3}=p_{4}=p_{3}^{0}=p_{4}^{0}$ \\ 
$T=T_{3}=T_{4}=T_{i}=T_{3}^{0}=T_{4}^{0}$%
\end{tabular}%
\end{equation}%
Cette hypoth\`{e}se suppose entre autres que l'\'{e}quilibre thermique s'%
\'{e}tablit suffisamment rapidement. Si les d\'{e}formations du solide sont
petites, on obtient les relations de Gibbs, d'Euler et de Gibbs-Duhem du mat%
\'{e}riau global :%
\begin{equation}
\begin{tabular}{l}
$T\frac{D}{Dt}\left( \frac{S}{\rho }\right) =\frac{D}{Dt}\left( \frac{U}{%
\rho }\right) +p\frac{D}{Dt}\left( \frac{1}{\rho }\right) -\frac{1}{\rho }%
\utilde{\sigma _{3}^{e}}^{s}:\frac{d_{3}\utilde{\varepsilon _{3}}^{s}}{dt}%
-\sum\limits_{1,2}\mu _{k}\frac{\rho _{4}}{\rho }\frac{d_{4}}{dt}\left( 
\frac{\rho _{k}}{\rho _{4}}\right) $ \\ 
$p=TS-U+\sum\limits_{k=1,2,3}\mu _{k}\rho _{k}$ \\ 
$\frac{dp}{dt}=S\frac{dT}{dt}+\sum\limits_{k=1,2,3}\rho _{k}\frac{d\mu _{k}}{%
dt}-\utilde{\sigma ^{e}}^{s}:\utilde{grad}\overrightarrow{V}$%
\end{tabular}%
\end{equation}%
La relation de Gibbs peut \'{e}galement s'\'{e}crire dans le r\'{e}f\'{e}%
rentiel barycentrique :%
\begin{equation}
T\frac{d}{dt}\left( \frac{S}{\rho }\right) =\frac{d}{dt}\left( \frac{U}{\rho 
}\right) +p\frac{d}{dt}\left( \frac{1}{\rho }\right)
-\sum\limits_{k=1,2,3}\mu _{k}\frac{d}{dt}\left( \frac{\rho _{k}}{\rho }%
\right) -\frac{1}{\rho }\utilde{\sigma ^{e}}^{s}:\utilde{grad}%
\overrightarrow{V}  \label{Gibbs}
\end{equation}

\subsection{Forces et flux gÈnÈralisÈs}
\medskip
Le tenseur des contraintes est l'addition du tenseur des contraintes d'\'{e}%
quilibre $\utilde{\sigma ^{e}}$ et du tenseur des contraintes visqueuses $%
\utilde{\sigma ^{v}}$, cette seconde partie \'{e}tant nulle \`{a} l'\'{e}%
quilibre : 
\begin{equation}
\utilde{\sigma }=\utilde{\sigma ^{e}}+\utilde{\sigma ^{v}}=-p\utilde{1}+%
\utilde{\sigma ^{e}}^{s}+\utilde{\sigma ^{v}}
\end{equation}%
En combinant les bilans d'\'{e}nergie interne (\ref{U}) et d'entropie (\ref%
{S}) avec la relation de Gibbs (\ref{Gibbs}), on peut identifier la cr\'{e}%
ation et le flux d'entropie : 
\begin{equation}
\begin{tabular}{l}
$s=\frac{\utilde{\sigma ^{v}}:\utilde{grad}\overrightarrow{V}}{T}+%
\frac{\overrightarrow{i^{\prime }}\cdot \overrightarrow{E}}{T}-%
\overrightarrow{Q^{\prime }}\cdot \frac{\overrightarrow{grad}T}{T^{2}}+\sum%
\limits_{k=1,2,3}\rho _{k}\left( \overrightarrow{V}-\overrightarrow{V_{k}}%
\right) \cdot \overrightarrow{grad}\frac{\mu _{k}}{T}$ \\ 
$\overrightarrow{\Sigma ^{\prime }}=\frac{\overrightarrow{Q^{\prime }}}{T}%
+\sum\limits_{k=1,2,3}\frac{\mu _{k}\rho _{k}}{T}\left( \overrightarrow{V}-%
\overrightarrow{V_{k}}\right) $%
\end{tabular}%
\end{equation}%
Introduisons dans $s$ les flux de diffusion de masse des cations dans la
solution $\overrightarrow{J_{1}}$ et de la solution dans le solide $%
\overrightarrow{J_{4}}$ :%
\begin{equation}
\overrightarrow{J_{1}}=\rho _{1}\left( \overrightarrow{V_{1}}-%
\overrightarrow{V_{2}}\right) \qquad \qquad \qquad \overrightarrow{J_{4}}%
=\rho _{4}\left( \overrightarrow{V_{4}}-\overrightarrow{V_{3}}\right)
\end{equation}%
Ces deux flux sont lin\'{e}airement ind\'{e}pendants. On identifie 
un flux scalaire, trois flux vectoriels et un flux tensoriel
ainsi que les forces g\'{e}n\'{e}ralis\'{e}es associ\'{e}es :%
\begin{equation}
\begin{tabular}{|l|l|}
\hline
Flux & Forces \\ \hline
$\frac{1}{3}tr\utilde{\sigma ^{v}}$ & $\frac{1}{T}div\overrightarrow{V}$ \\ 
\hline
$\overrightarrow{Q^{\prime }}$ & $\overrightarrow{grad}\frac{1}{T}$ \\ \hline
$\overrightarrow{J_{1}}$ & $\frac{\rho _{2}}{\rho _{4}}\left[ \frac{1}{T}%
Z_{1}\overrightarrow{E}-\overrightarrow{grad}\frac{\mu _{1}}{T}+%
\overrightarrow{grad}\frac{\mu _{2}}{T}\right] $ \\ \hline
$\overrightarrow{J_{4}}$ & $\frac{\rho _{3}}{\rho }\left[ \frac{1}{T}\left( 
\frac{\rho _{1}}{\rho _{4}}Z_{1}-Z_{3}\right) \overrightarrow{E}-\frac{\rho
_{1}}{\rho _{4}}\overrightarrow{grad}\frac{\mu _{1}}{T}-\frac{\rho _{2}}{%
\rho _{4}}\overrightarrow{grad}\frac{\mu _{2}}{T}+\overrightarrow{grad}\frac{%
\mu _{3}}{T}\right] $ \\ \hline
$\utilde{\sigma ^{v}}^{s}$ & $\frac{1}{T}\utilde{grad}\overrightarrow{V}^{s}$
\\ \hline
\end{tabular}
\label{IdentFlux}
\end{equation}

\section{Equations constitutives}

\subsection{Loi rhÈologique}
\medskip
Le milieu est isotrope. D'apr\`{e}s le principe de Curie, il ne peut donc
pas y avoir de couplage entre des forces et des flux d'ordres tensoriels diff%
\'{e}rents \cite{De Groot}. Compte tenu de la sym\'{e}trie du tenseur $%
\utilde{\sigma^{v}}^{s}$, les flux scalaire et tensoriel d'ordre 2 s'\'{e}%
crivent donc :%
\begin{equation}
\begin{tabular}{l}
$\frac{1}{3}tr\left( \utilde{\sigma ^{v}}\right) =\frac{L_{1}}{T}div%
\overrightarrow{V}$ \\ 
$\utilde{\sigma ^{v}}^{s}=\frac{L_{2}}{T}\utilde{grad}\overrightarrow{V}^{s}$%
\end{tabular}%
\end{equation}%
o\`{u} $L_{1}$ et $L_{2}$ sont deux coefficients ph\'{e}nom\'{e}nologiques
scalaires. $\utilde{grad}\overrightarrow{V}^{s}$ d\'{e}signe la partie sym%
\'{e}trique sans trace du tenseur $\utilde{grad}\overrightarrow{V}$. En
supposant que le tenseur des contraintes d'\'{e}quilibre v\'{e}rifie la loi
de Hooke, on obtient :%
\begin{equation}
\utilde{\sigma }=\lambda \left( tr\utilde{\varepsilon }\right) \utilde{1}+2G%
\utilde{\varepsilon }+\frac{L_{1}^{\prime }}{T}\left( tr\overset{\bullet }{%
\utilde{\varepsilon }}\right) \utilde{1}+\frac{L_{2}}{T}\overset{\bullet }{%
\utilde{\varepsilon }}
\end{equation}%
o\`{u} $L_{1}^{\prime }=L_{1}-\frac{L_{2}}{3}$; $\lambda $ est le premier
coefficient de Lam\'{e}, $G$ le module de cisaillement et $%
\utilde{\varepsilon }$ le tenseur des d\'{e}formations du mat\'{e}riau
complet. Si la phase liquide est un fluide newtonien et stok\'{e}sien, la pression v%
\'{e}rifie :%
\begin{equation}
p=-\frac{1}{3}tr\left( \utilde{\sigma ^{e}}\right) =\left( \lambda +\frac{2}{%
3}G\right) tr\utilde{\varepsilon }
\end{equation}%
La loi rh\'{e}ologique obtenue s'identifie avec un mod\`{e}le de Kelvin
-Voigt de coefficients visco\'{e}lastiques $\lambda _{v}$ et $\mu _{v}$ :%
\begin{equation}
\lambda _{v}\equiv \frac{L_{1}^{\prime }}{T}\qquad \qquad 2\mu _{v}\equiv 
\frac{L_{2}}{T}
\end{equation}%

Les diff\'{e}rents coefficients peuvent \^{e}tre estim\'{e}s en se basant
sur le cas du Nafion satur\'{e} en eau, bien document\'{e} dans la litt\'{e}%
rature. On obtient les ordres de grandeur suivants \cite{Silberstein2010,Satterfield2009,Bauer} :%
\begin{equation}
G\sim 4.5\;10^{7}\;Pa\qquad \lambda \sim 3\;10^{8}\;Pa\qquad E\sim
1.3\;10^{8}\;Pa\qquad \nu \sim 0.435
\end{equation}%
o\`{u} $E$ d\'{e}signe le module d'Young et $\nu $ le coefficient de
Poisson. Le temps de relaxation en traction est de l'ordre de $\mathit{15}\;%
\mathit{s}$ \cite{Silberstein2010,Silberstein2008,Silberstein2011} et il est
proche de celui en relaxation en cisaillement \cite{Combette}. Les
coefficients visco\'{e}lastiques peuvent \^{e}tre d\'{e}duits d'essais de
traction \cite{Silberstein2010,Satterfield2009,Silberstein2011} et des temps
de relaxation :%
\begin{equation}
\lambda _{v}\sim \;7\;10^{8}\;Pa\;s\qquad \mu _{v}\sim 10^{8}\;Pa\;s
\end{equation}

\noindent On en d\'{e}duit :%
\begin{equation}
L_{1}^{^{\prime }}\sim 2.1\;10^{11}\;Pa\;s\;K\qquad \qquad L_{2}\sim
6\;10^{10}\;Pa\;s\;K
\end{equation}

\noindent Ces coefficients d\'{e}pendent fortement de la temp\'{e}rature,
qui est ici assez proche de la temp\'{e}rature de transition vitreuse.

\subsection{Loi de Fourier gÈnÈralisÈe}
\medskip
La premiËre \'{e}quation constitutive vectorielle s'\'{e}crit :%
\begin{equation}
\begin{tabular}{l}
$\overrightarrow{Q^{\prime }}=L_{3}\overrightarrow{grad}\frac{1}{T}+L_{4}%
\frac{\rho _{2}}{\rho _{4}}\left[ \frac{1}{T}Z_{1}\overrightarrow{E}-%
\overrightarrow{grad}\frac{\mu _{1}}{T}+\overrightarrow{grad}\frac{\mu _{2}}{%
T}\right] $ \\ 
$\quad +L_{5}\frac{\rho _{3}}{\rho }\left[ \frac{1}{T}\left( \frac{\rho _{1}%
}{\rho _{4}}Z_{1}-Z_{3}\right) \overrightarrow{E}-\frac{\rho _{1}}{\rho _{4}}%
\overrightarrow{grad}\frac{\mu _{1}}{T}-\frac{\rho _{2}}{\rho _{4}}%
\overrightarrow{grad}\frac{\mu _{2}}{T}+\overrightarrow{grad}\frac{\mu _{3}}{%
T}\right] $%
\end{tabular}%
\end{equation}%

La phase liquide est une solution dilu\'{e}e d'\'{e}lectrolyte fort. D'aprËs \cite{Diu} :%
\begin{equation}
\begin{tabular}{l}
$\mu _{1}\left( T,p,x\right) \simeq \mu _{1}^{0}\left( T,p\right) +\frac{RT}{%
M_{1}}\ln \left( C\frac{M_{2}}{\rho _{2}^{0}}\right) $ \\ 
$\mu _{2}\left( T,p,x\right) \simeq \mu _{2}^{0}\left( T,p\right) -\frac{RT}{%
\rho _{2}^{0}}C$ \\ 
$\mu _{3}\left( T,p,x\right) =\mu _{3}^{0}\left( T\right) $%
\end{tabular}%
\end{equation}%
o\`{u} $\mu _{2}^{0}$ et $\mu _{3}^{0}$ d\'{e}signent les potentiels
chimiques du solide et du solvant pur et o\`{u} $\mu _{1}^{0}$ d\'{e}pend du
solvant et du solut\'{e}. Ces expressions permettent de calculer $%
\overrightarrow{grad}\mu _{k}$ compte tenu des relations de Gibbs-Duhem des
phases solide et liquide.

\section{Discussion}

\subsection{PropriÈtÈs physiques du Nafion}
\medskip
Nous allons approximer les deux autres relations vectorielles en nous limitant 
au cas isotherme et en nous basant sur le cas du Nafion 
\cite{nemat2000,Heitner-Wirguin,Gebel,Cappadonia,Choi}. On obtient les valeurs
suivantes :%
\begin{equation*}
\begin{tabular}{|ll|l|l|l|}
\hline
&  & Cations & Solvant & Solide \\ \hline
$M_{k}$ & $\left( kg\;mol^{-1}\right) $ & $\mathit{10}^{\mathit{-2}}$ & $%
\mathit{18}\;\mathit{10}^{\mathit{-3}}$ & $\mathit{10}^{\mathit{2}}$ - $%
\mathit{10}^{\mathit{3}}$ \\ \hline
$\rho _{k}^{0}$ & $\left( kg\;m^{-3}\right) $ &  & $\mathit{10}^{\mathit{3}}$
& $\mathit{2.1}\;\mathit{10}^{\mathit{3}}$ \\ \hline
$v_{k}$ & $\left( m^{3}\;mol^{-1}\right) $ & $\frac{M_{1}}{\rho _{4}^{0}}%
\sim \mathit{10}^{\mathit{-5}}$ & $\mathit{18}\;\mathit{10}^{\mathit{-6}}$ & 
\\ \hline
$\rho _{k}$ & $\left( kg\;m^{-3}\right) $ & $\mathit{14}$ & $\mathit{0.35\;10%
}^{3}$ & $\mathit{1.4\;10}^{3}$ \\ \hline
$Z_{k}$ & $\left( C\;kg^{-1}\right) $ & $\mathit{10}^{\mathit{7}}$ & $%
\mathit{0}$ & $\mathit{9\;10}^{\mathit{4}}$ \\ \hline
\end{tabular}%
\end{equation*}%
o\`{u} $v_{k}$ est le volume molaire partiel du constituant $k$. $C\sim
4\;10^{3}\;mol\;m^{-3}$ et la viscositÈ dynamique de l'eau est 
$\eta _{2}=\mathit{10}^{\mathit{-3}}\;Pa\;s$. La
masse \'{e}quivalente du polym\`{e}re ou masse de polym\`{e}re par mole de
sites ioniques vaut $M_{eq}$ $\sim \mathit{1.1}\;kg\;eq^{-1}$. On prendra en
outre $\phi _{4}\sim \mathit{35\%}$, $T=300\;K$ et $\left\Vert 
\overrightarrow{E}\right\Vert \sim 10^{4}\;V\;m^{-1}$. On en d\'{e}duit $%
\rho \sim 1.8\;10^{3}\;kg\;m^{-3}$. En premi\`{e}re approximation, on peut
donc consid\'{e}rer que :%
\begin{equation}
Z_{1}>>Z_{3}\qquad \qquad \rho \sim \rho _{2}\sim \rho _{3}>>\rho _{1}\qquad
\qquad \rho _{1}Z_{1}\sim \rho _{4}Z_{3}
\end{equation}

\subsection{Equation de Nernst-Planck}
\medskip
Les coefficients ph\'{e}nom\'{e}nologiques diagonaux sont g\'{e}n\'{e}%
ralement grands devant les coefficients non diagonaux :%
\begin{equation}
L_{6}\gtrsim L_{7}
\end{equation}%
On en d\'{e}duit :%
\begin{equation}
\overrightarrow{J_{1}}\simeq \frac{L_{6}Z_{1}}{T}\overrightarrow{E}+\frac{1}{%
T}\left[ \frac{v_{2}}{M_{2}}\left( L_{6}-\frac{\rho _{3}L_{7}}{\rho }\right)
-\frac{v_{1}L_{6}}{M_{1}}\right] \overrightarrow{grad}p-\frac{RL_{6}}{M_{1}C}%
\overrightarrow{grad}C
\end{equation}%
expression que l'on peut identifier avec la relation de Nernst-Planck \cite%
{Lakshmi} :%
\begin{equation}
\overrightarrow{V_{1}}=-\frac{D}{C}\left[ \overrightarrow{grad}C-\frac{%
Z_{1}M_{1}C}{RT}\overrightarrow{E}+\frac{Cv_{1}}{RT}\left( 1-\frac{M_{1}}{%
M_{2}}\frac{v_{2}}{v_{1}}\right) \overrightarrow{grad}p\right] +%
\overrightarrow{V_{2}}
\end{equation}%
$R=8,314\;J\;K^{-1}$ est la constante universelle des gaz parfaits. 
$D=\frac{RL_{6}}{M_{1}\rho _{1}}\sim \mathit{2}\;\mathit{10}^{\mathit{-9}}%
m^{2}\;s^{-1}$ est le coefficient de diffusion de masse
des cations \cite{Zawodsinski}. On en d\'{e}duit :%
\begin{equation}
L_{6}\sim 3,5\;10^{-11}\;kg\;K\;s\;m^{-3}\qquad \qquad L_{7}<<L_{6}
\end{equation}%
On peut estimer les diff\'{e}rents termes de l'\'{e}quation.
L'ordre de grandeur du gradient de concentration des cations peut \^{e}tre d%
\'{e}duit de \cite{nemat2000} et \cite{nemat2002}. Le gradient de pression est 
de l'ordre de $10^{9}\;Pa\;m^{-1}$. D'o\`{u} :%
\begin{equation}
\begin{tabular}{ll}
$\left\Vert \overrightarrow{grad}C\right\Vert $ & $\lesssim
10^{8}\;mol\;m^{-4}$ \\ 
$\frac{M_{1}C}{RT}Z_{1}\left\Vert \overrightarrow{E}\right\Vert $ & $\sim
1.6\;10^{9}\;mol\;m^{-4}$ \\ 
$\frac{Cv_{1}}{RT}\left( 1-\frac{M_{1}}{M_{2}}\frac{v_{2}}{v_{1}}\right)
\left\Vert \overrightarrow{grad}p\right\Vert $ & $\sim
1.1\;10^{3}\;mol\;m^{-4}$%
\end{tabular}%
\end{equation}%
Le champ \'{e}lectrique et la diffusion de masse sont donc les principaux
responsables du d\'{e}placement des cations; le gradient de pression a un
effet n\'{e}gligeable.

\subsection{Loi de Darcy gÈnÈralisÈe}
\medskip
Compte tenu des hypoth\`{e}ses faites, le flux $\overrightarrow{J_{4}}$ peut
s'\'{e}crire : 
\begin{equation}
\begin{tabular}{l}
$\overrightarrow{J_{4}}\simeq \frac{1}{T}\left[ L_{7}Z_{1}+L_{8}\frac{\rho
_{3}}{\rho }\left( \frac{\rho _{1}}{\rho _{4}}Z_{1}-Z_{3}\right) \right] 
\overrightarrow{E}-\frac{R}{M_{1}C}L_{7}\overrightarrow{grad}C$ \\ 
$\quad +\frac{1}{T}\left[ L_{7}\left( \frac{v_{2}}{M_{2}}-\frac{v_{1}}{M_{1}}%
\right) -\frac{\rho _{3}\phi _{4}}{\rho \rho _{4}}L_{8}\right] 
\overrightarrow{grad}p$%
\end{tabular}%
\end{equation}%
En identifiant le terme de pression \`{a} la loi de Darcy, on obtient :%
\begin{equation}
L_{8}\sim \frac{KT}{\eta _{2}\phi _{4}^{2}}\frac{\rho _{2}^{2}\rho }{\rho
_{3}}\sim 3.8\;10^{-5}\;kg\;s\;K\;m^{-3}>>L_{7}
\end{equation}%
o\`{u} $K$ d\'{e}signe la perm\'{e}abilit\'{e} intrins\`{e}que de la phase
solide et $\eta _{2}$ la viscosit\'{e} dynamique du solvant; compte tenu de
la taille des pores ($\mathit{100}$\ \AA ; \cite{Gebel}), $K\sim \mathit{10}%
^{\mathit{-16}}\;m^{2}$. Les ordres de grandeur des diff\'{e}rents termes de
l'\'{e}quation sont les suivants :%
\begin{equation}
\begin{tabular}{ll}
$\frac{K}{\eta _{2}\phi _{4}}\left\Vert \overrightarrow{grad}p\right\Vert $
& $\sim 2.8\;10^{-4}\;m\;s^{-1}$ \\ 
$\frac{K\rho _{2}^{0}}{\eta _{2}\phi _{4}}\left( \frac{\rho _{1}}{\rho _{4}}%
Z_{1}-Z_{3}\right) \left\Vert \overrightarrow{E}\right\Vert $ & $\sim
1.1\;m\;s^{-1}$ \\ 
$\frac{R}{M_{1}C\rho _{4}}L^{8}\left\Vert \overrightarrow{grad}C\right\Vert $
& $<<2\;10^{-6}\;m\;s^{-1}$%
\end{tabular}%
\end{equation}%
La relation ph\'{e}nom\'{e}nologique obtenue s'identifie avec une loi
de Darcy g\'{e}n\'{e}ralis\'{e}e :%
\begin{equation}
\overrightarrow{V_{4}}-\overrightarrow{V_{3}}\simeq -\frac{K}{\eta _{2}\phi
_{4}}\left[ \overrightarrow{grad}p-\rho _{4}^{0}\left( Z_{4}-Z_{3}\right) 
\overrightarrow{E}\right]
\end{equation}%
Le second terme de cette expression traduit le mouvement de la solution sous
l'action du champ \'{e}lectrique; il s'agit donc d'un terme d'osmose. La
distribution des cations est tr\`{e}s h\'{e}t\'{e}rog\`{e}ne \cite{nemat2002}
: ils s'accumulent pr\`{e}s de l'\'{e}lectrode n\'{e}gative o\`{u} l'on peut 
\'{e}crire $Z_{4}>>Z_{3}$. L'\'{e}quation constitutive obtenue co\"{\i}ncide
alors avec celle de M.A. Biot \cite{Biot}. Au voisinage de l'\'{e}%
lectrode n\'{e}gative, $Z_{4}<<Z_{3}$ et l'on retrouve le r\'{e}sultat de
Grimshaw et al \cite{nemat2000,Grimshaw}. Au centre de la lame, les deux
termes ont le m\^{e}me ordre de grandeur.

\section{Conclusion}
\medskip
Nous avons mod\'{e}lis\'{e} le comportement d'un polym\`{e}re \'{e}%
lectroactif ionique satur\'{e} d'eau de type Nafion. La pr\'{e}sence d'eau
provoque une dissociation quasi totale du polym\`{e}re et la lib\'{e}ration
de cations de petite taille. Nous avons repr\'{e}sent\'{e} ce milieu par la
superposition de trois syst\`{e}mes ayant des champs de vitesses diff\'{e}%
rents : les cations, le solvant et le solide assimil\'{e} \`{a} un milieu
poreux d\'{e}formable. Nous avons \'{e}crit les \'{e}quations de bilan de la
masse, de la quantit\'{e} de mouvement, de l'\'{e}nergie et de l'entropie et
les \'{e}quations de Maxwell pour chaque phase (solide et liquide) \`{a} l'%
\'{e}chelle microscopique. Un processus de moyenne nous a permis d'en d\'{e}%
duire les \'{e}quations relatives au milieu complet \`{a} l'\'{e}chelle
macroscopique. L'\'{e}criture des relations thermodynamiques nous permet
d'obtenir la fonction de dissipation du mat\'{e}riau.

Nous en avons d\'{e}duit ses lois de comportement : une loi rh\'{e}ologique
de type Kelvin-Voigt, une loi de Fourier et une loi de Darcy g\'{e}n\'{e}%
ralis\'{e}es et une \'{e}quation de Nernst-Planck. Nous avons fait une \'{e}%
valuation des coefficients ph\'{e}nom\'{e}nologiques et des diff\'{e}rents
termes de ces relations en fonction des param\`{e}tres physiques des
constituants.

Nous envisageons maintenant de comparer les r\'{e}sultats fournis par ce mod%
\`{e}le aux donn\'{e}es exp\'{e}rimentales publi\'{e}es dans la litt\'{e}%
rature. Une piste d'am\'{e}lioration pourrait \^{e}tre de remplacer la loi rh%
\'{e}ologique par un mod\`{e}le de Zener, mieux adapt\'{e} au comportement
visco\'{e}lastique des polym\`{e}res.

\section{Notations}
\medskip
Les indices $k=1,2,3,4,i$ d\'{e}signent respectivement les cations, le
solvant, le solide, la solution et l'interface. Les quantit\'{e}s non indic%
\'{e}es sont relatives au mat\'{e}riau complet. L'exposant $^{0}$ indique
une quantit\'{e} \`{a} l'\'{e}chelle microscopique, l'absence d'exposant
correspond \`{a} l'\'{e}chelle macroscopique. Les quantit\'{e}s
microscopiques sont rapport\'{e}es au volume de la phase correspondante, les
quantit\'{e}s macroscopique au volume du mat\'{e}riau complet. L'exposant $%
^{s}$ d\'{e}signe la partie sym\'{e}trique sans trace d'un tenseur du second
ordre, et $^{T}$ sa transpos\'{e}e.\newline
\newline
\noindent$C$ : concentration molaire en cations (relative \`{a} la phase
liquide);\newline
\noindent$D$ : coefficient de diffusion de masse des cations dans la phase
liquide;\newline
\noindent$\overrightarrow{D}$ ($\overrightarrow{D_{k}^{0}}$) : induction \'{e}lectrique;\newline
\noindent$\overrightarrow{E}$ ($\overrightarrow{E_{k}^{0}}$) : champ \'{e}lectrique;\newline
\noindent$E$ ($E_{c}$, $E_{p}$, $U$, $U_{k}^{0}$) : \'{e}nergie volumique totale (cin\'{e}tique, potentielle, interne);\newline
\noindent$G$, $\lambda $, $E$, $\nu $ : coefficients \'{e}lastiques;\newline
\noindent$\overrightarrow{I}\ $($\overrightarrow{I_{k}}$) : densit\'{e} volumique de courant;\newline
\noindent$\overrightarrow{i}\ $($\overrightarrow{i^{\prime }}$) : courant \'{e}lectrique de diffusion;\newline
\noindent$\overrightarrow{J_{k}}$ : flux de diffusion massique;\newline
\noindent$K$ : perm\'{e}abilit\'{e} intrins\`{e}que de la phase solide;\newline
\noindent$L_{i},L_{i}^{\prime }$ : coefficients ph\'{e}nom\'{e}nologiques;\newline
\noindent$M_{k}$ : masse molaire du constituant $k$;\newline
\noindent$\overrightarrow{n_{k}}$ : normale sortante de la phase $k$;\newline
\noindent$p$ ($p_{k}$, $p_{k}^{0}$) : pression;\newline
\noindent$\overrightarrow{Q}$ ($\overrightarrow{Q^{\prime }}$) : flux de chaleur;\newline
\noindent$s$ : production volumique d'entropie;\newline
\noindent$S$ ($S_{k}$) : entropie volumique;\newline
\noindent$T$ ($T_{k}$, $T_{k}^{0}$) : temp\'{e}rature absolue;\newline
\noindent$v_{k}$ : volume molaire partiel du constituant $k$ (relatif \`{a} la phase liquide);\newline
\noindent$\overrightarrow{V}$ ($\overrightarrow{V_{k}}$, $\overrightarrow{V_{k}^{0}}$) : vitesse;\newline
\noindent$Z$ ($Z_{k}$) : charge \'{e}lectrique massique;\newline
\noindent$\varepsilon $ ($\varepsilon _{k}^{0}$) : permittivit\'{e} di\'{e}lectrique;\newline
\noindent$\utilde{\varepsilon }$ ($\utilde{\varepsilon _{k}}$, $\utilde{\varepsilon _{k}^{0}}$) : tenseur
des d\'{e}formations;\newline
\noindent$\eta _{2}$ : viscosit\'{e} dynamique de l'eau;\newline
\noindent$\lambda _{v}$\textit{, }$\mu _{v}$\textit{\ }: coefficients viscoelastiques;\newline
\noindent$\mu _{k}$ ($\mu _{k}^{0}$) : potentiel chimique massique;\newline
\noindent$\rho $ ($\rho _{k}$, $\rho _{k}^{\prime }$, $\rho _{k}^{0}$) : masse
volumique;\newline
\noindent$\utilde{\sigma }$ ($\utilde{\sigma ^{v}}$, $\utilde{\sigma ^{e}}$, $%
\utilde{\sigma_{k}}$, $\utilde{\sigma_{k}^{0e}}$) : tenseur des contraintes 
totales (dynamiques, d'\'{e}quilibre);\newline
\noindent$\overrightarrow{\Sigma }$ ($\overrightarrow{\Sigma ^{\prime }}$) :
flux d'entropie par conduction;\newline
\noindent$\phi _{k}$ : fraction volumique de la phase $k$;\newline
\noindent$\chi _{k}$ : fonction de pr\'{e}sence de la phase $k$ ;


%

\end{document}